# The relation between migration and FDI in the OECD from a complex network perspective

Antonios Garas

*Chair of Systems Design ETH Zurich, Weinbergstrasse 56/58, CH-8092 Zürich*

*Email: agaras@ethz.ch*

Athanasios Lapatinas[1]

*Department of Economics, University of Ioannina, P.O. Box 1186, 45110, Ioannina, Greece*

*Email: alapatin@cc.uoi.gr*

Konstantinos Poulios

*Department of Economics, University of Ioannina, P.O. Box 1186, 45110, Ioannina, Greece*

*Email: poulios.konstantinos@yahoo.gr*

**Abstract**

We explore the relationship between human migration and OECD's Foreign Direct Investment (FDI) using a gravity equation enriched with variables that account for complex-network effects. Based on a panel data analysis, we find a strong positive correlation between the migration network and the FDI network, which can be mostly explained by countries' economic/demographic sizes and geographical distance. We highlight the existence of a stronger positive FDI relationship in pairs of countries that are more central in the migration network. Both intensive and extensive forms of centrality are FDI enhancing. Illuminating this result, we show that bilateral FDI between any two countries is further affected positively by the complex web of 'third party' corridors/migration stocks of the international migration network. Our findings are consistent whether we consider bilateral FDI and bilateral migration figures, or we focus on the outward FDI and the respective inward migration of the OECD countries.

---

[1] Corresponding author.

# 1 Introduction

During the past decades there was a huge flow of people, capital and knowledge around the world, which can be mainly attributed to human migration across borders and Foreign Direct Investment (FDI). And while throughout the literature it has been of major interest to explore the mechanisms of globalization, little attention has been paid in answering the question of whether immigration is related to FDI. As we will discuss in the next section, only a small part of literature has explored so far possible links between FDI and migration, suggesting that cross-border capital flows are affected by bilateral migration. As argued, the migration of people brings to the destination country factors of production, like capital and labor, but also a social network connected to immigrants' origin country. These social networks may lower potential barriers to international investment, as immigrants possess crucial information about the structure of the local market, the preferences, as well as the business ethics and the commercial codes. This knowledge can be proved invaluable for overcoming many informational and contractual barriers, leading to stimulated investment activities across national boundaries.

Whereas previous studies have focused on migrants within a single country, this work analyzes migration and FDI between many countries. The paper brings together a wide range of bilateral migration, FDI positions, geopolitical, demographic, economic, and socioeconomic data for 34 OECD and 185 partner countries, covering three five-year periods that span from 1995 to 2010, and investigates the correlation between the international migration network and the international FDI network. More precisely, the current paper contributes to the existing literature by investigating the topological properties of the OECD's bilateral FDI network and the OECD's bilateral migration network. Considering a complex-network perspective we study the correlation patterns of the two networks. We follow Fagiolo and Mastrorillo (2014), as well as Sgrignoli et al. (2015), who explore similar issues using a trade perspective instead of FDI. We fit gravity models of bilateral FDI where we introduce centrality measures of the international migration network as explanatory variables. Our main hypothesis is that bilateral FDI is positively affected by the corresponding pair of countries' centrality in the migration network.

In our analysis, we first scout for similarities between the two networks by comparing links, topological structures and node statistics, finding strong correlations which can be mostly explained by countries' economic, demographic and geographical differences. Next, we specifically explore whether pairs of countries that are more central in the migration network foreign invest more. Thus, we add migration-network variables in our gravity regression equations of bilateral FDI stocks/positions in order to control for countries' network-centralization and the intensity of 'third-party' migration origins. Confirming the results of the previous literature, we find that the networks of international migration and bilateral FDI are strongly and positively correlated, and such correlation can be mostly explained by countries' economic/demographic sizes and geographical distance. More interestingly, we find that centrality in the international migration network boosts bilateral FDI between any two countries, and this result is further enhanced by countries' relative embeddedness in the complex web of corridors making up the network of international human migration. Overall, we find that bilateral FDI between any two countries is not only affected by the presence of migrants from either countries but also by the number of their total inward-migration links. In other words, FDI between any two countries increases due to their binary connectivity in the international migration network, and in turn, due to a higher number of inward corridors coming from common 'third party' migration origins.[2]

We suggest that this *indirect network effect* in our results may be driven by learning processes of new investment preferences by immigrants from 'third party' origins and/or by the presence in both countries of second-generation immigrants belonging to the same ethnic group. Moreover, the positive effect of the international migration network on bilateral FDI may be due to linkages created not only between the origin and destination of migration, but also among countries that are the destinations of migration flows from third countries. More immigrants coming from 'third party' origins may imply more openness and inclusive institutions in both economies, which may in turn foster learning processes about investment patterns and therefore more bilateral capital exchanges (Fagiolo and Mastrorillo, 2014).

In this paper, similar to Fagiolo and Mastrorillo (2014), we further test the extent to which binary (extensive) vs. weighted (intensive) inward country centrality explain bilateral FDI. Hence, we also take into account weighted centrality indicators to further understand the role played by the stock of immigrants (intensive migration margins). We find that both forms of migration separately increase bilateral FDI. Hence, expanding inward migration corridors and attracting a larger number of immigrants separately seem important in increasing FDI. However, when their effects are jointly considered, intensive inward centrality outweighs its extensive counterpart, thus indicating that the stock of immigrants is more FDI enhancing than the number of inward migration corridors, confirming the analogous result found by Fagiolo and Mastrorillo (2014) for bilateral trade.

---

[2] We use only inward corridors as we believe that outward channels are less relevant in explaining bilateral FDI.

Moreover, in order to shed more light on our results and to control that the co-movement of FDI and migration flows is not driven by the identification issue that capital investments and migration between two countries may be caused by demand shocks we also explore the relationship between country $i$'s inward migration from country $j$, and country $i$'s outward FDI to country $j$. We find that outward FDI position is again positively affected by inward migration stocks and that the result is again enhanced by the FDI-parent country's in-degree position in the migration network. Thus, our results are consistent whether we consider bilateral FDI and migration figures, or we focus on the outward FDI and inward migration of the OECD countries.

The remainder of the paper is structured as follows. The next section briefly reviews the related literature. Thereafter, section 3 describes our dataset and visualizes the two networks, while section 4 presents the topological properties of the two networks. The empirical results are discussed in Section 5. Finally, section 6 concludes our analysis.

## 2  Related Literature

The literature linking FDI and migration is relatively scarce and usually refers either to within a particular country's migrants affecting bilateral investment with their country-of-birth, or to the migrants from a particular country living in a number of other countries affecting capital flows between those other countries. The gravity equation model for bilateral trade flows is increasingly used to analyze FDI (Wei, 2000; Razin and Sadka, 2007; Blonigen et al. 2007). Bergstrand and Egger (2007) and Head and Ries (2008) develop the leading theoretical models that provide theoretical micro-foundations for adopting gravity equations for the analysis of FDI.

Clemens and Williamson (2000) find that, historically, British foreign capital that flowed into countries also attracted a large number of migrants. In the same line, Barry (2002) uses aggregated data to show that migration has an impact on inflows of FDI, while Gao (2003) finds that FDI into China is positively related to the share of the Chinese population in the FDI-parent country. Hunt (2004) finds that migration within Germany often takes the form of same-employer migration and Tong (2005) shows that the number of ethnic Chinese in both the FDI-parent and the FDI-host country is positively correlated with the cumulative amount of their reciprocal FDI. Kugler and Rapoport (2005) find a positive impact of the change in immigrants from a particular origin-country into the US on outward FDI of US firms into this country. Buch et al. (2006) show that there are higher stocks of inward FDI in German states hosting a large foreign population from the same country of origin, while Kugler and Rapoport (2007) demonstrate that migration and FDI inflows are negatively correlated contemporaneously but migration is associated with an increase in future FDI. Bhattacharya and Groznik (2008) find that US outward foreign investment in a country is higher the higher the income of the immigrant group from that country living in the US is. Furthermore, Ligthart and Singer (2009) investigate the role of immigrants in Dutch outward FDI and find that they facilitate outward FDI to their countries of origin.

More recently, Leblang (2010) tests the hypothesis of whether diaspora networks influence cross-border investment by reducing transaction and information costs. He uses dyadic cross-sectional data for portfolio and FDI and he finds a substantively and statistically significant effect of diaspora networks on global investment. Moreover, Javorcik et al. (2011) investigate the link between the presence of migrants in the US and US FDI in the migrants' countries of origin, addressing potential endogeneity of migration with respect to FDI by employing the instrumental variables approach. They conclude that the presence of migrants in the US increases the volume of US FDI in their country of origin. Flisi and Murat (2011) focus on the relation between bilateral FDI and skilled and unskilled immigrants and they observe that FDI of UK, Germany and France is prompted by the ties of skilled immigrants, while the FDI of Italy and Spain is only influenced by their respective diasporas. Finally, Foad (2012) improves identification issues by looking at the US-regional distributions of FDI and immigration. Using a unique measure of immigrant network size for each US-state, he finds that immigration tends to lead FDI.

Regarding the complex-network perspective of our approach, to the best of our knowledge although the topological properties of the international migration network and its evolution over time has been explored (Fagiolo and Mastrorillo, 2014; Sgrignoli et al., 2015), an investigation of the relation between the international migration network and FDI is missing. The current paper not only explores the topological properties of the OECD's outward FDI network but it does so by jointly investigating FDI and migration as dependent phenomena i.e. as if they were two fully connected layers of the directed-weighted multi-graph where nodes are world countries and links represent their macroeconomic interaction channels (Schweitzer et al., 2009).[3]

## 3    Data and Network Visualizations

This section outlines the data sources on migrant and FDI figures, as well as other explanatory variables. A bilateral FDI and bilateral migration panel was constructed for 34 OECD countries and up to 185 partner countries for three time periods, totaling 7,625 observations, including FDI -zeros.[4] Following the relative literature we utilize FDI stocks, as FDI flows are very volatile and therefore harder to model. Thus, outward (OECD to partner country) and inward (partner country to OECD) FDI positions were sourced from the *OECD International Direct Investment Database*, and are presented in US dollars.[5] The OECD countries included in our dataset hosted 71% of global inward FDI and were the source of 87% of global outward FDI in 2000 (UNCTAD, 2006).

Furthermore, we retrieved origin-destination (bilateral) migration data, for all the countries in the FDI dataset, from Abel and Sander (2014). The authors quantify international migration flows at the country level and present mid-year to mid-year data for four five-year periods, totally spanning from 1990 to 2010. The estimates capture the number of people who change their country of residence during these periods. In this work, we utilize three time periods, overall spanning from 1995 to 2010, due to FDI data availability. The fact that Abel and Sander (2014) consider migration flows until the half of the last year of every five-year period, allows us to use the five-year migration flow estimate as a lagged determinant of the FDI position documented at the end of the five-year period. Thus, the structure of Abel and Sander's (2014) dataset drives the specifics of the econometric model used in our analysis, resulting in a pooled panel dataset, which consists of three time periods.

---

[3] See also Battiston et al. (2007) for a complex-network based analysis of inter-regional investment stocks within Europe.

[4] We handle zero-FDI observations by using three alternative approaches which provide qualitatively similar empirical results: (a) Poisson estimation (b) Poisson pseudo-maximum likelihood estimation and (c) negative binomial estimation.

[5] Details on the data may be found at stats.oecd.org.

Bilateral FDI stocks can be described by a gravity equation that relates the log of bilateral investment to the logged economic sizes of OECD and partner economies and the logged distance between them. Thus, in our gravity equation, we include logged values of the GDP per capita and population figures for every country pair, taken from the *World Development Indicators* published by the *World Bank*. We complete the gravity equation by including the log distance between an OECD-partner pair, measured as the longitudinal distance in kilometers between the biggest cities in the two countries, weighted by the share of the city in the overall country's population. The distance dataset is retrieved by *CEPII* (see www.cepii.fr).

In the regression analysis that follows, we also use bilateral country geopolitical and socioeconomic data, in order to deal with potential identification issues like cultural similarity, an issue not properly addressed by the relevant literature. Cultural similarities could render the exchange of migrants and FDI between two partner countries more attractive. The variables used to control for this issue are time invariant and are included in *CEPII's* geodist dataset. The dataset contains information about whether the two countries have ever had a colonial link, share a language, or used to be part of the same country.

We use bilateral FDI positions and bilateral migration data to build two weighted-undirected networks wherein for each network, between any two nodes, there is one weighted-undirected link. This link describes total bilateral capital movements and bilateral migration respectively. The generic element of the international FDI network (IFDIN) records the log of total bilateral FDI stocks/positions. I.e., $TFDI_{ij}$ is the stock of FDI that country $i$ owns in country $j$ plus the stock of FDI that country $j$ owns in country $i$ ($TFDI_{ij} = FDI_{ij} + FDI_{ji}$), where the index $i$ denotes the 34 OECD countries and the index $j$ denotes 185 countries for which we have FDI data for the years 2000, 2005 and 2010. On the other hand, the generic element of the international migration network (IMN) represents the log of total bilateral migrants, $tmig_{ij} = m_{ij} + m_{ji}$, for every five-year period. Accordingly, we define the binary projection of the two networks through their adjacency matrices, where their generic elements are equal to one if the corresponding entry in the weighted version is strictly positive.

Similar to Abel and Sander (2014), Figure 1 shows a circular plot visualizing the top 5% of the networks' link weights for both the migration network (panel a) and the bilateral FDI network (panel b) in years 2000, 2005 and 2010. For the migration network, these years refer to the end of every five-year period. The size of the circular segments representing individual countries are scaled proportionally to the strength of the corresponding country in the respective network.

In Figure 1a, we see the pronounced role of USA, which is the most important node with respect to FDI stock exchange. Other pronounced nodes include large EU economies such as Great Britain, France, Germany and the Netherlands. It is therefore evident that the most important capital movements emanate from prosperous countries as the highest volume of FDI stock is transferred mostly among OECD countries. In comparison, the fraction of non-OECD countries participating in the top 5% of FDI stock exchanges is negligible, and it is mostly between these countries and the USA. This behavior is consistent across time, even though the network increases in density. This means that more capital is exchanged over time, but the lion's share circulates mostly among developed countries.

As expected, Figure 1b shows that the presence of low-income countries is more pronounced in the migration flow network. However, even in this network, the largest flows occur among OECD countries. Once more the country with the most prominent role is USA followed by large EU countries. It is worth noting, though, that USA is more involved in global migrant flows, while EU migration is dominated by internal migration among member states. Similar to the FDI network, these observations are stable over time.

**Figure 1.** Circular plot visualizing the FDI Network (a) and the Migration Network (b) in years 2000, 2005 and 2010.

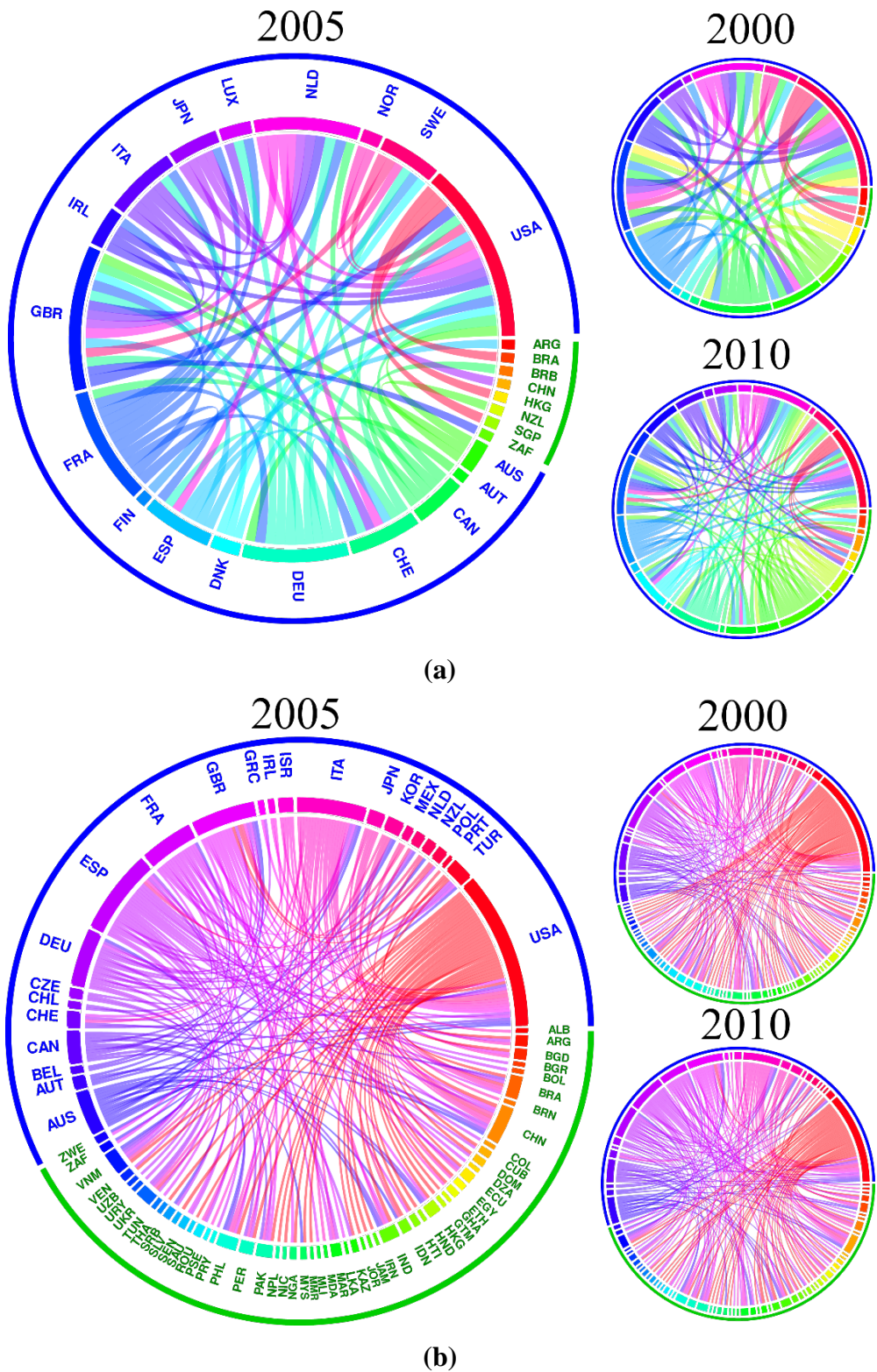

(a)

(b)

*Notes:* Only the top 5% of the networks' link weights are drawn. The external large blue segment groups together OECD countries, while the external green segment non-OECD countries. Each internal circular segment represents an individual country, and its size is proportional to the node strength of this country in the respective network, while the width of the flow represents the link weight.

# 4     Descriptive Analysis of the two Networks

Following Newman (2010), we compute basic descriptive statistics of the two networks across time, as shown in Table 1. More precisely, the list of calculated descriptive statistics includes the average node degree, the network density, the number of (the weakly) connected components, the network diameter and the average path length (APL). These are accompanied by topological measures showing structural correlations such as the assortativity coefficient and the network transitivity (or clustering coefficient). Comparing the topological properties of the two networks across time we find that the migration network is much denser that the FDI network. However, the FDI network has almost doubled its density from 2000 to 2010, while the density of the migration network remained at the same level. Both networks are connected and have relatively small diameter and APL, which indicates the presence of small world properties. In this respect, the migration network features a more pronounced small-world property with a smaller average path length than in the FDI network. However, the APL value of the migration network remained almost stable over the three examined periods, while on the other hand the APL of the FDI network has decreased. Furthermore, the global clustering coefficient (transitivity) is always larger in the migration network, even though over the years a measurable increasing trend for the FDI network is observed. This implies that countries in the migration network have a higher tendency to form clusters i.e. if there are migration flows between countries *(A,B)*, and *(B,C)*, then there is a high probability for migration flows between countries *(A,C)*. In addition, the (strong) negative assortativity coefficients we find for both networks indicate that, as it is expected, capital and migration relationships occur mostly between countries with different degree centralities.

**Table 1.** Descriptive Network Statistics

|  | 2000 | | 2005 | | 2010 | |
| --- | --- | --- | --- | --- | --- | --- |
|  | IFDIN | IMN | IFDIN | IMN | IFDIN | IMN |
| No. Nodes | 169 | 185 | 180 | 185 | 183 | 185 |
| Mean degree | 9.480 | 47.892 | 13.733 | 47.524 | 18.197 | 47.286 |
| Density | 0.056 | 0.260 | 0.077 | 0.258 | 0.099 | 0.257 |
| APL (undirected) | 2.013 | 1.740 | 1.973 | 1.742 | 1.901 | 1.743 |
| No. WCC | 1 | 1 | 1 | 1 | 1 | 1 |
| Diameter | 4 | 3 | 4 | 2 | 3 | 3 |
| Assortativity | -0.518 | -0.755 | -0.563 | -0.757 | -0.567 | -0.767 |
| Transitivity | 0.246 | 0.410 | 0.258 | 0.411 | 0.315 | 0.398 |

*Notes:* IFDIN: International FDI Network; IMN: International Migration Network; APL: Average Path Length; WCC: Weakly connected components.

Further network similarities are studied by exploring link weights' correlation. In Figure 2 we provide a scatter plot of the link weights in the FDI network against the link weights in the migration network (log scale) for the year 2005. Note that a stronger link-weight in the FDI network is typically associated with a stronger migration link-weight, and that this positive relation is explained by countries' economic sizes, demographic sizes and geographic distances (note that in Figure 2, markers' size is proportional to the logged product of country populations divided by country distance. Colors scale -from lighter to darker- is from lower to higher values of the logged product of countries' per capita GDPs divided by country distance), stimulating the adoption of a gravity-like equation in the next section.

**Figure 2.** International migration network (IMN) versus international FDI network (IFDIN) link weights in year 2005

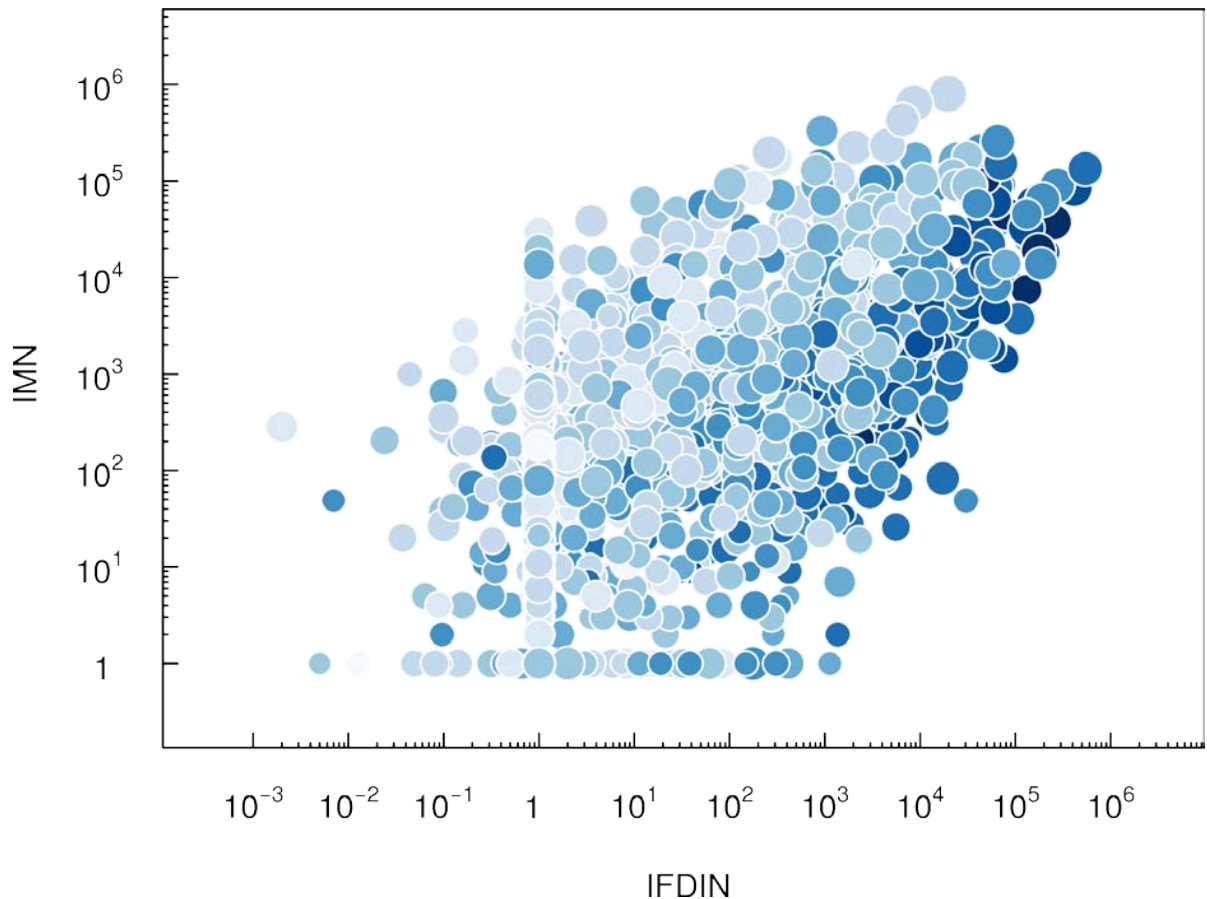

*Notes:* Logarithmic scale. Markers' size is proportional to the logged product of country populations divided by country distance. Colors scale (from lighter to darker) is from lower to higher values of the logged product of countries' per capita GDPs divided by country distance.

Finalizing our descriptive network analysis, we compute the correlations between the two networks' node-statistics for the year 2005. Panel (a) in Figure 3 indicates that node strengths are positively and linearly correlated in the two networks. This finding can be explained by countries' economic and demographic differences (note again that in Figure 3, markers' size is proportional to the logged product of country populations divided by country distance. Colors scale -from lighter to darker- is from lower to higher values of the logged product of countries' per capita GDPs divided by country distance), and it means that the more a country foreign invests the larger the immigrant stocks it holds. Panel (b) indicates that Average Nearest Neighbor Strength (ANNS) is positively correlated in the two networks implying that if a country is FDI connected to a country that in turn is highly FDI connected with third-party countries, it will also be migration-connected to countries that hold a lot of migration stocks. Again, demographic and economic country characteristics are associated with the above finding but now in a different manner, namely, countries with larger levels of ANNS are smaller and poorer. The above findings motivate the inclusion of network variables, as additional controls, in next section's gravity equation.

**Figure 3.** Correlation of node network statistics between international migration network (IMN) and international FDI network (IFDIN) in year 2005.

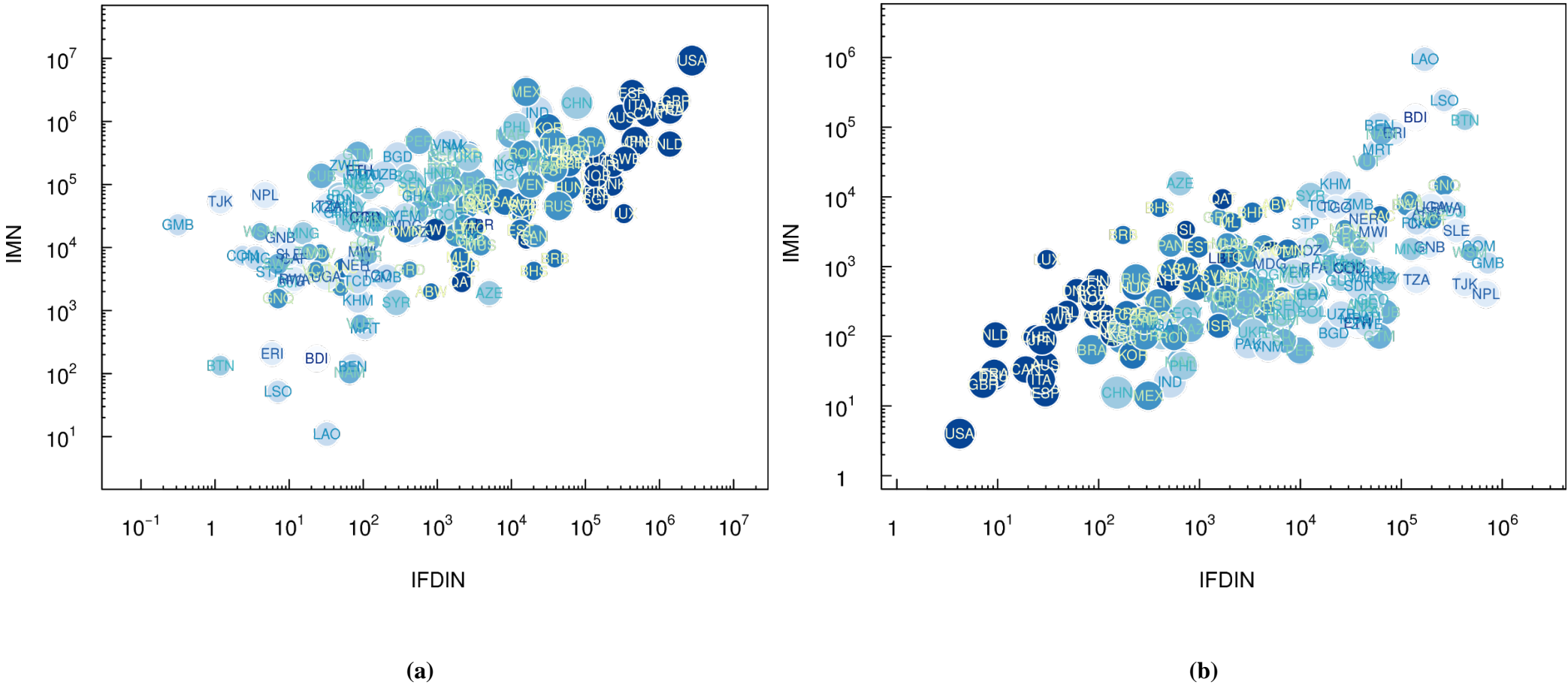

(a)　　　　　　　　　　　　　　　　　　(b)

*Notes:* Panel (a): Total strength; Panel (b): Average Nearest-Neighbor Strength (ANNS); Markers' size is proportional to logs of population; Colors scale (from lighter to darker) is from lower to higher (logged) values of GDP per capita.

# 5  Empirical Analysis

In this section, we combine the panel dataset on international migration flows from Abel and Sander (2014) with data on FDI in order to provide an empirical study of the relationship between the two networks. We use a gravity model approach, while controlling for network effects. We first test if the networks of international migration and FDI are correlated. Then we ask whether pairs of countries that are more central in the migration network exchange more capital. Finally, we investigate whether bilateral FDI is further affected by the complex web of 'third party' corridors of the international migration network (Table 2). Moreover, in order to control that the co-movement of FDI and migration flows is not driven by the identification issue that capital investments and migration into a FDI-host country may be caused by a demand shock in the FDI-parent country, we additionally estimate only the one direction of migration -the opposite one to FDI-, namely, the effect of inward migration on outward FDI (Table 3).

Our empirical analysis starts with the recognition of the much-discussed issue -in the relevant literature- of zero-FDI values. Previous studies have used Ordinary Least Squares (OLS) to investigate the empirical relationship between FDI and its determinants (Chakrabati, 2001). However, this method may be inappropriate for analyzing the count number of FDI, as it assumes that the dependent variable follows a normal distribution. In the case of FDI, (a) the dependent variable cannot be negative and (b) there are many zero counts due to some countries not receiving any FDI at all; hence, the dependent variable is not normally distributed. To deal with count data we estimate the following baseline equation, using the framework of Poisson regression model with robust standard errors, country $i$, country $j$ fixed effects as well as time fixed effects[6]:

$$\log(TFDI_{ijt}) = \alpha_1 \log(pop_{it-1}) + \alpha_2 \log(pop_{jt-1}) + a_3 \log(pcgdp_{it-1}) \\ + a_4 \log(pcgdp_{jt-1}) + a_5 \log(distwces_{ij}) + a_6 language_{ij} + a_7 colony_{ij} \\ + a_8 col45_{ij} + a_9 smctry_{ij} + \beta \log(tmig_{ijt-1}) + \gamma_1 ind^b_{ijt-1} + \gamma'_1 ind^w_{ijt-1} \\ + \gamma_2 overin^b_{ijt-1} + \gamma'_2 overin^w_{ijt-1} + c_1 + c_i + c_j + c_t + \varepsilon_{ijt}$$

---

[6] The basic idea of the Poisson regression was outlined by Coleman (1964, 378-379). An early example of Poisson regression was Cochran (1940). See McNeil (1996), Long (1997), Selvin (2004) Johnson et al. (2005), Selvin (2011), Long and Freese (2014) and Cameron and Trivedi (2013) for textbook treatments and Allison (2009) for an extensive discussion on these models.

(1)

where $TFDI_{ijt}$ is the total bilateral FDI stocks at time t (years 2000, 2005, 2010) defined as $TFDI_{ij} = FDI_{ij} + FDI_{ji}$; $pop_{it-1}$, $pop_{jt-1}$ and $pcgdp_{it-1}$, $pcgdp_{jt-1}$ are countries' population and GDP per capita respectively, at time $t-1$ (years 1999, 2004, 2009). $distwces_{ij}$ is longitudinal distance in kilometers between the biggest cities in countries $i$ and $j$, weighted by the share of the city in the overall country's population. Moreover, the dummy variables introduced indicate whether the two countries share a language spoken by at least 9% of the population in both countries ($language_{ij}$), have ever had a colonial link ($colony_{ij}$) or a colonial relationship after 1945 ($col45_{ij}$) and used to be part of the same country ($smctry_{ij}$). $tmig_{ijt-1}$ is the total bilateral migration stock at time $t-1$, defined as $tmig_{ijt-1} = m_{ijt-1} + m_{jit-1}$.[7] $ind^b_{ijt-1}$ (resp. $ind^w_{ijt-1}$) is a binary (resp. weighted) centrality indicator measuring the total in-degree centralization, constructed as the logged sum of inward (resp. weighted) links of country $i$ and the inward (resp. weighted) links of country $j$. $overin^b_{ijt-1}$ (resp. $overin^w_{ijt-1}$) captures the effect of third-country common inward migration channels and it is constructed as the logged sum of inward (resp. weighted) links of country $i$ and inward (resp. weighted) links of country $j$ originated from common third country $k$. Thus, the variable $overin^b_{ijt-1}$ (resp. $overin^w_{ijt-1}$) sums up only the commonly-shared (overlapping) inward migration (resp. weighted) links originated from third countries $k$, while $ind^b_{ijt-1}$ (resp. $ind^w_{ijt-1}$) sums up the total inward migration (weighted) links (overlapping and non-overlapping). $c_1$ is a constant, $c_i, c_j, c_t$ are OECD country, partner country and time fixed effects, respectively. Finally, $\varepsilon_{ij}$ is the error term. Notice that we lag all the time varying independent variables, in order to address possible issues of reverse causality.

We further examine the relationship between the two networks by replacing in equation (1) the total bilateral FDI positions ($TFDI_{ijt}$) with the outward FDI position of country $i$ in country $j$ and the total bilateral migration stock ($tmig_{ij}$) with the stock of migrants originated from country $j$ (FDI-host country) and present in country $i$ (FDI-parent country), demonstrating that our empirical results on capital and people co-movements between two countries are not driven by a demand shock. $ind^b_{it-1}$ (resp. $ind^w_{it-1}$) is a binary (resp. weighted) centrality indicator constructed as the log in-degree centralization of the FDI-parent country i.e. log value of total inward (resp. weighted) links of country $i$. Thus, our second specification is:

$$\begin{aligned}\log(FDI_{ijt}) = &\tilde{\alpha}_1\log(pop_{it-1}) + \tilde{\alpha}_2\log(pop_{jt-1}) + \tilde{\alpha}_3\log(pcgdp_{it-1}) + \tilde{\alpha}_4\log(pcgdp_{jt-1})\\ &+ \tilde{\alpha}_5\log(distwces_{ij}) + \tilde{\alpha}_6 language_{ij} + \tilde{\alpha}_7 colony_{ij} + \tilde{\alpha}_8 col45_{ij}\\ &+ \tilde{\alpha}_9 smctry_{ij} + \tilde{\beta}\log(m_{jit-1}) + \tilde{\gamma} ind^b_{it-1} + \tilde{\gamma}' ind^w_{it-1} + \tilde{c}_1 + \tilde{c}_i + \tilde{c}_j + \tilde{c}_t\\ &+ \tilde{\varepsilon}_{ijt}\end{aligned}$$

(2)

---
[7] Abel and Sander's, 2014, bilateral migration flows for the five-year periods: mid1995-mid2000, mid2000-mid2005 and mid2005-mid2010.

**Table 2.** Poisson Regression Results (Total FDI and Total Migration)

|  | (1) | (2) | (3) | (4) | (5) | (6) |
|---|---|---|---|---|---|---|
| $\log(pop_i)$ | -5.361*** | -5.027*** | -5.524*** | -5.528*** | -5.285*** | -5.376*** |
|  | (1.411) | (1.411) | (1.404) | (1.410) | (1.412) | (1.416) |
| $\log(pop_j)$ | -0.099 | -0.160 | -0.707** | -0.706** | -0.139 | -0.300 |
|  | (0.322) | (0.322) | (0.318) | (0.318) | (0.214) | (0.321) |
| $\log(pcgdp_i)$ | -0.572*** | -0.693*** | -0.613*** | -0.612*** | -0.598*** | -0.595*** |
|  | (0.214) | (0.215) | (0.211) | (0.214) | (0.214) | (0.213) |
| $\log(pcgdp_j)$ | 0.272*** | 0.198* | 0.270*** | 0.271*** | 0.228** | 0.171* |
|  | (0.102) | (0.104) | (0.099) | (0.102) | (0.105) | (0.101) |
| $\log(distwces_{ij})$ | -0.316*** | -0.326*** | -0.300*** | -0.300*** | -0.306*** | -0.290*** |
|  | (0.041) | (0.041) | (0.041) | (0.041) | (0.041) | (0.041) |
| $language_{ij}$ | 0.066* | 0.066* | 0.079** | 0.079** | 0.065* | 0.066* |
|  | (0.038) | (0.038) | (0.037) | (0.037) | (0.038) | (0.038) |
| $colony_{ij}$ | 0.170*** | 0.168*** | 0.190*** | 0.190*** | 0.168*** | 0.172*** |
|  | (0.047) | (0.047) | (0.044) | (0.044) | (0.047) | (0.046) |
| $col45_{ij}$ | 0.244*** | 0.242*** | 0.158** | 0.158** | 0.242*** | 0.208** |
|  | (0.077) | (0.076) | (0.075) | (0.075) | (0.077) | (0.076) |
| $smctry_{ij}$ | 0.256*** | 0.256** | 0.314*** | 0.314*** | 0.258*** | 0.280*** |
|  | (0.100) | (0.100) | (0.101) | (0.101) | (0.100) | (0.099) |
| $\log(tmig_{ij})$ | 0.128*** | 0.126*** | 0.127*** | 0.127*** | 0.132*** | 0.134*** |
|  | (0.014) | (0.014) | (0.014) | (0.014) | (0.014) | (0.014) |
| $ind^b_{ij}$ |  | 0.674*** |  | -0.007 |  |  |
|  |  | (0.183) |  | (0.179) |  |  |
| $ind^w_{ij}$ |  |  | 0.684*** | 0.684*** |  |  |
|  |  |  | (0.044) | (0.045) |  |  |
| $overin^b_{ij}$ |  |  |  |  | 0.106** |  |
|  |  |  |  |  | (0.050) |  |
| $overin^w_{ij}$ |  |  |  |  |  | 0.193*** |
|  |  |  |  |  |  | (0.030) |
| Fixed Effects | | | Country $i$ / country $j$ / Time | | | |
| Pseudo $R^2$ | 0.435 | 0.436 | 0.444 | 0.444 | 0.436 | 0.438 |
| Wald $X^2$ | 8827.650 | 8820.880 | 8797.950 | 8801.260 | 8825.720 | 8901.490 |
|  | [0.000] | [0.000] | [0.000] | [0.000] | [0.000] | [0.000] |
| Log Pseudolikelihood | -16654.228 | -16640.930 | -16402.637 | -16402.636 | -16649.300 | -16560.512 |
| No of Observations | 7625 | 7625 | 7625 | 7625 | 7625 | 7625 |

*Notes*: Dependent Variable: $\log(TFDI_{ijt})$. Independent Variables: see text. Poisson regressions are estimated with country $i$, country $j$ and time fixed effects. Numbers in parentheses are robust standard errors; p-values in brackets. The symbols *, ** and *** reveal statistical significance at 10%, 5% and 1% respectively.

**Table 3.** Poisson Regression Results (Outward FDI and Inward Migration)

|  | (1) | (2) | (3) | (4) |
|---|---|---|---|---|
| $\log(pop_i)$ | -3.553*** | -3.516*** | -4.059*** | -4.108*** |
|  | (1.059) | (1.061) | (1.069) | (1.074) |
| $\log(pop_j)$ | -0.021 | -0.023 | -0.016 | -0.014 |
|  | (0.275) | (0.274) | (0.273) | (0.273) |
| $\log(pcgdp_i)$ | -0.151 | -0.168 | -0.225 | -0.209 |
|  | (0.189) | (0.196) | (0.189) | (0.195) |
| $\log(pcgdp_j)$ | 0.351*** | 0.349*** | 0.339*** | 0.340*** |
|  | (0.090) | (0.090) | (0.089) | (0.089) |
| $\log(distwces_{ij})$ | -0.396*** | -0.397*** | -0.402*** | -0.401*** |
|  | (0.031) | (0.031) | (0.031) | (0.031) |
| $language_{ij}$ | 0.128*** | 0.128*** | 0.131*** | 0.131*** |
|  | (0.031) | (0.031) | (0.031) | (0.031) |
| $colony_{ij}$ | 0.107*** | 0.107*** | 0.106*** | 0.106*** |
|  | (0.039) | (0.039) | (0.039) | (0.039) |
| $col45_{ij}$ | 0.361*** | 0.361*** | 0.366*** | 0.366*** |
|  | (0.064) | (0.064) | (0.064) | (0.064) |
| $smctry_{ij}$ | 0.271*** | 0.271*** | 0.266*** | 0.269*** |
|  | (0.082) | (0.082) | (0.082) | (0.082) |
| $\log(m_{ji})$ | 0.068*** | 0.068*** | 0.063*** | 0.063*** |
|  | (0.007) | (0.007) | (0.007) | (0.007) |
| $ind_i^b$ |  | 0.069 |  | -0.073 |
|  |  | (0.156) |  | (0.157) |
| $ind_i^w$ |  |  | 0.253*** | 0.258*** |
|  |  |  | (0.059) | (0.059) |
| Fixed Effects | Country $i$ / country $j$ / Time | | | |
| $Pseudo\ R^2$ | 0.421 | 0.421 | 0.422 | 0.422 |
| $Wald\ X^2$ | 10874.940 | 10879.050 | 10852.860 | 10856.780 |
|  | [0.000] | [0.000] | [0.000] | [0.000] |
| Log Pseudolikelihood | -22307.427 | -22307.293 | -22293.201 | -22293.058 |
| No of Observations | 9938 | 9938 | 9938 | 9938 |

*Notes*: Dependent Variable: $\log(FDI_{ijt})$. Independent Variables: see text. Poisson regressions are estimated with country $i$, country $j$ and time fixed effects. Numbers in parentheses are robust standard errors; p-values in brackets. The symbols *, ** and *** reveal statistical significance at 10%, 5% and 1% respectively.

Note that from equation (1) the bilateral migration stocks in a pair of countries is found to be related with increased bilateral foreign direct investments. Respectively, from equation (2) we find that the migration stocks originated from country $j$ and present in country $i$ are associated with increased capital investments of country $i$ in country $j$. Thus, we confirm the related literature's finding that migrant diasporas indeed contribute to attracting investment. The impact of the control variables is in most cases strong, significant and signed as expected. Variables like the populations of countries in both specifications and the GDP per capita of country $i$ in the first specification, initially raise concerns about their lack of statistical significance and their signs. However, a thorough examination reveals that when we do not include country fixed effects, the variables are statistically significant and positively signed. Thus, a possible explanation is that the inclusion of fixed effects absorbs the effects of these variables, while the negative coefficients pick up population growth and per capita GDP, which may be correlated with poor institutional quality, causing a negative coefficient.[8]

A supplementary contribution of this paper is the inclusion of the network variables as additional control variables in the gravity equations. Our results show that bilateral foreign direct investments between any two countries is not only affected by the presence of migrants from either countries, but also by their relative embeddedness in the complex web of corridors making up the network of international human migration. The positive sign and the statistical significance of the network variables in the first econometric specification strongly suggest that bilateral FDI increases, the more the two countries under consideration are inward central in the international migration network. Interestingly, inward migrants coming from common 'third party' corridors are FDI enhancing. We argue that our results may be driven either by learning processes of new investment preferences by migrants whose origins are shared by the two countries or by the presence in both countries of second-generation migrants belonging to the same ethnic group. Our findings indicate that migration networks are conductive to bilateral investment because they create linkages not only between pairs of countries that are the origin and destination of migration, but also among countries that are the destinations of migration flows originated from common third countries (Fagiolo and Mastrorillo, 2014).

The estimation of equation (2), when only the effect of inward migration on outward FDI is considered, reveals the consistency of our results and shows that the network weighted variable has again a positive and statistically significant effect on outward FDI but its binary counterpart is statistically insignificant. This implies that the more inward-migration stocks the FDI-parent country holds, the higher its foreign capital investments to the migration-origin country.

---

[8] The estimated results without the country fixed effects are available upon request.

We also test whether our findings remain intact when considering the intensive form of centrality into the IMN i.e. weighted inward country centrality (Table 2). As expected, both extensive and intensive inward centralities separately boost bilateral FDI (columns 2 and 3; columns 5 and 6). Column 4 also depicts the case where extensive and intensive inward centralities are jointly considered in the Poisson regression. It is clear that the coefficient of the extensive form loses its importance in explaining bilateral FDI, indicating that intensive inward centrality has no effect on bilateral FDI. Therefore, confirming the analogous result found by Fagiolo and Mastrorillo (2014) for bilateral trade, it seems that intensive centrality in the IMN outweighs extensive centrality, i.e. bilateral FDI seems to be boosted more by the number of immigrants than by the number of inward corridors held by any two countries in the IMN.

As a robustness check of our results, we also consider a Poisson estimation by pseudo-maximum likelihood (PML). It differs from the Poisson regression because it uses the method of Santos Silva and Tenreyro (2010) to identify and drop regressors that may cause the non-existence of the (pseudo) maximum likelihood estimates. In our case, the results from this specification are identical to those of the Poisson regressions (Table 2). We also consider a negative binomial regression (Hausman et al., 1984) as an additional robustness check. Table 4 presents the results, which remain qualitatively intact to this alternative technique. Hence, it seems that our findings remain robust to alternative estimation models.

**Table 4.** Robustness checks: Negative Binomial (NB)

| | (1) | (2) | (3) | (4) | (5) | (6) |
|---|---|---|---|---|---|---|
| $log(pop_i)$ | -7.343*** | -6.852*** | -7.283*** | -7.320*** | -7.216*** | -7.037*** |
| | (1.820) | (1.817) | (1.815) | (1.821) | (1.819) | (1.811) |
| $log(pop_j)$ | -0.148 | -0.213 | -0.937** | -0.936** | -0.196 | -0.493 |
| | (0.424) | (0.424) | (0.423) | (0.423) | (0.425) | (0.424) |
| $log(pcgdp_i)$ | -0.784*** | -0.953*** | -0.850*** | -0.840*** | -0.818*** | -0.845*** |
| | (0.284) | (0.286) | (0.278) | (0.283) | (0.284) | (0.277) |
| $log(pcgdp_j)$ | 0.354*** | 0.262** | 0.347*** | 0.353*** | 0.303** | 0.231* |
| | (0.130) | (0.133) | (0.126) | (0.129) | (0.133) | (0.128) |
| $log(distwces_{ij})$ | -0.302*** | -0.315*** | -0.290*** | -0.299*** | -0.291*** | -0.261*** |
| | (0.050) | (0.050) | (0.050) | (0.050) | (0.050) | (0.049) |
| $language_{ij}$ | 0.090* | 0.089* | 0.103** | 0.103** | 0.089* | 0.091* |
| | (0.047) | (0.047) | (0.046) | (0.046) | (0.047) | (0.047) |
| $colony_{ij}$ | 0.184*** | 0.184*** | 0.218*** | 0.218*** | 0.182*** | 0.192*** |
| | (0.056) | (0.056) | (0.053) | (0.053) | (0.056) | (0.054) |
| $col45_{ij}$ | 0.192** | 0.187** | 0.071 | 0.071 | 0.193** | 0.133 |
| | (0.096) | (0.096) | (0.093) | (0.093) | (0.096) | (0.094) |
| $smctry_{ij}$ | 0.267** | 0.269** | 0.373*** | 0.374*** | 0.269** | 0.319*** |
| | (0.125) | (0.125) | (0.129) | (0.129) | (0.125) | (0.124) |
| $log(tmig_{ij})$ | 0.162*** | 0.159*** | 0.154*** | 0.154*** | 0.165*** | 0.169*** |
| | (0.018) | (0.018) | (0.018) | (0.018) | (0.018) | (0.018) |
| $ind_{ij}^b$ | | 0.815*** | | -0.056 | | |
| | | (0.233) | | (0.227) | | |
| $ind_{ij}^w$ | | | 0.875*** | 0.878*** | | |
| | | | (0.057) | (0.058) | | |
| $overin_{ij}^b$ | | | | | 0.123* | |
| | | | | | (0.067) | |
| $overin_{ij}^w$ | | | | | | 0.315*** |
| | | | | | | (0.053) |
| Fixed Effects | Country $i$ / country $j$ / Time | | | | | |
| $Pseudo\ R^2$ | 0.230 | 0.231 | 0.236 | 0.236 | 0.230 | 0.233 |
| $Wald\ X^2$ | 9259.990 | 9260.710 | 9213.650 | 9221.930 | 9257.880 | 9295.02 |
| | [0.000] | [0.000] | [0.000] | [0.000] | [0.000] | [0.000] |
| Log Pseudolikelihood | -13648.292 | -13642.922 | -13537.954 | -13537.930 | -13646.644 | -13601.423 |
| No of Observations | 7625 | 7625 | 7625 | 7625 | 7625 | 7625 |

*Notes*: Dependent Variable: $log(TFDI_{ijt})$. Independent Variables: see text. Negative binomial regressions are estimated with country $i$, country $j$ and time fixed effects. Numbers in parentheses are robust standard errors; p-values in brackets. The symbols *, ** and *** reveal statistical significance at 10%, 5% and 1% respectively.

# 6  Conclusions

Throughout the world, economies are becoming rapidly integrated and the level of dependence between them increases substantially. Globalization has led to a rapid growth in the flow of factors of production across borders. From 1980 to 2010, there has been an increase of about 65 million in the foreign population in the OECD countries, while the volume of FDI grew four times as fast as world output during the same period. The international flow of people and capital are important features of this integrated global economy and taken together, the international investment channels and the migration corridors constitute a convoluted and complicated web of relationships among countries.

This paper has explored the properties and the link between migration and FDI, using a gravity model enriched with complex-network effects. Diasporas in the OECD attract FDI to their origin countries and this result can be mostly explained by countries' economic, demographic and geographic characteristics. Our main findings though suggest that bilateral FDI increases the more inward central in the migration network pairs of countries are. Moreover, migrants originated from overlapping 'third party' countries can be FDI enhancing. We have also found that outward FDI is positively associated to inward migration: migrants in country $j$ originated from country $i$ attract FDI in their origin country $i$ from destination country $j$. Interestingly, our results indicate that the larger the diversity of migration channels and the stock of immigrants from 'third-party' origins towards any two countries that are FDI connected, the higher the stock of foreign capital in the FDI-host country originated from the FDI-parent country. Our findings remain robust to alternative count data estimation techniques.

With our work we add to the present literature an exploratory analysis, which highlights the existence of previously unexplored sources of influence in the relation between FDI and migration. Furthermore, we do believe that there is space for additional improvement of this approach. Particularly, we suggest an examination of a wide set of immigrant characteristics which, along with network variables could provide further insight on the relationship between human migration and FDI. Higher frequency of the migration data should also provide a considerable improvement for future studies since so far migration datasets are based on censuses conducted every ten years, while FDI data are updated annually, creating a frequency mismatch. Finally, the focus on the current paper was placed on direct investment that generally builds on a wide network of economic agents, requiring a long-run focus on the characteristics of the host country. Thus, the examination of how human migration affects short-term portfolio investment flows could provide us with interesting results.

**Acknowledgements:** Antonios Garas acknowledges support by the EU-FET project MULTIPLEX 317532